\documentclass[a4paper,11pt]{article}
\pdfoutput=1 % if your are submitting a pdflatex (i.e. if you have
\usepackage{jheppub} % for details on the use of the package, please
\usepackage{bbm}

\usepackage[T1]{fontenc} % if needed

\title{\boldmath One-loop effective potential in Scherk--Schwarz compactifications of pure $d=5$ supergravities}

\author{Gianguido Dall'Agata}
\author{and Fabio Zwirner}
\affiliation{Dipartimento di Fisica e Astronomia `G.~Galilei', Universit\`a di Padova 
\\ 
and INFN, Sezione di Padova, Via Marzolo 8, I-35131 Padova, Italy}
\emailAdd{gianguido.dallagata@pd.infn.it}
\emailAdd{fabio.zwirner@pd.infn.it}
\abstract{
We perform a systematic analysis of the one-loop effective potential of pure $d=5$ supergravities, with supersymmetry fully broken by a Scherk--Schwarz compactification on the circle, as a function of the radial modulus.
We discuss the precise correspondence between the effective potential $V_1$ in the full compactified theory and  its counterpart $V_{1,red}$ in the reduced theory.
We confirm that $V_1$ is finite for any $N>0$, in contrast to $V_{1,red}$.
We find that for broken $N=8$ supergravity $V_1$ is negative definite even after accounting for the Kaluza--Klein states. 
We outline a program for future work where the study of a different kind of Scherk-Schwarz compactifications, still at the field theory level but with at least three extra dimensions, could lead to qualitatively new results.
}
\def\simlt{\mathrel{\lower2.5pt\vbox{\lineskip=0pt\baselineskip=0pt
           \hbox{$<$}\hbox{$\sim$}}}}
\def\simgt{\mathrel{\lower2.5pt\vbox{\lineskip=0pt\baselineskip=0pt
           \hbox{$>$}\hbox{$\sim$}}}}

\begin{document} 
\maketitle
\flushbottom

\section{Introduction} % (fold)
\label{sec:introduction}

Questions such as the approximate stability of a classical Minkowski background and the dynamical generation of hierarchies can be quantitatively addressed in higher-dimensional supergravity or superstring models where local supersymmetry is broken in the compactification, for example by the Scherk--Schwarz mechanism \cite{Scherk:1979zr}. 
The obvious first tool for such an exploration is  \cite{Coleman:1973jx,Weinberg:1973ua} the one-loop effective potential $V_1$, computed as a function of those background fields whose constant value is left undetermined at the classical level.

The corresponding reduced four-dimensional ($d=4$) effective supergravities all fall in the broad category of no-scale models \cite{Cremmer:1983bf}, in the sense that the scale of supersymmetry breaking on the classical Minkowski vacuum is controlled by one or more undetermined moduli fields. 
In these reduced theories, which contain only the finite number of degrees of freedom that would be massless in the limit of unbroken supersymmetry,  $V_{1,red}$ is controlled by the supertraces of the field-dependent mass matrix, and is in general divergent. 
The (field-independent) quartic divergence, which would be present in a generic $d=4$ theory, is proportional to $Str \, {\cal M}^0$, and is absent if the reduced theory has a spontaneously broken supersymmetry, so that the number of bosonic and fermionic degrees of freedom is the same, $n_B=n_F$.  
Reduced theories with spontaneously broken ${\cal N}=1$, $d=4$ supersymmetry have in general a quadratically divergent contribution to $V_{1,red}$ proportional to ${\rm Str} \, {\cal M}^2$, although in special cases one can achieve its automatic cancellation. 
Only in reduced theories with spontaneously broken ${\cal N}=4$, $d=4$ supergravity it is guaranteed that ${\rm Str} \, {\cal M}^2=0$ and the divergent contribution to $V_{1,red}$ is only logarithmically divergent and proportional to ${\rm Str} \, {\cal M}^4$. 
Reduced theories with spontaneously broken ${\cal N}=8$, $d=4$ supergravity have the special property that ${\rm Str} \, {\cal M}^2= {\rm Str} \, {\cal M}^4 = {\rm Str} \, {\cal M}^6=0$ and $V_{1,red}$ is finite \cite{Cremmer:1979uq,DallAgata:2012tne}, although it was found to be negative definite in all cases studied so far, so that no locally stable Minkowski or de Sitter vacuum has been found. This last result is somewhat disappointing, but as we will discuss now it cannot be considered to be the end of the story. 

Indeed, the behaviour of the reduced $d=4$ supergravity theories does not capture the physics of the full compactified theory, which in the case of supergravity contains also infinite towers of Kaluza--Klein modes, with supersymmetry-breaking mass splittings that depend on the background values of classically massless moduli fields, associated with geometrical properties of the compactification manifold. 
In addition, string models compactified on smooth manifolds contain additional winding modes, corresponding to strings wrapped around the compactification manifold, whose effects should exponentially decouple in the large-volume limit.  
String models with localized defects such as orbifold fixed points, D-branes, orientifold planes, etc, give rise to additional non-perturbative states in the spectrum and make the structure of $V_1$ more complicated to analyse and not always under full control. 

Some studies of $V_1$ in fully compactified theories have already been performed, in the context of rigid supersymmetry, of supergravity and of superstrings, often with emphasis on semi-realistic orbifold models, but a systematic understanding of the main physical effects that control the behaviour of $V_1$ and of their interplay is stil missing. 
We refer here, for example, to the effects of: the non-locality of supersymmetry breaking a la Scherk--Schwarz (and, more generally, by fluxes), which should guarantee \cite{Rohm:1983aq} the finiteness of $V_1$, and to the organization of the massive KK states in massive multiplets of ${\cal N} > 1$ supersymmetry, which would deserve a better understanding, going beyond the simplest compactifications.

A feature that could play a crucial role in determining the properties of $V_1$ is the difference between Scherk--Schwarz compactifications where the breaking of the higher-dimensional supersymmetry is explicit \cite{Scherk:1979zr}, through boundary conditions that are periodic up to an internal global R-symmetry, and those in which it is spontaneous \cite{Cremmer:1979uq}, through twists in the boundary conditions that are equivalent to constant VEVs of some internal components of the spin connection.
The latter case corresponds to a sort of `geometrical fluxes', and leads to the quantisation of the supersymmetry-breaking parameters when considering the full compactified theory (already at the supergravity level, without invoking string theory, as explained for example in \cite{Kaloper:1999yr}).
Finally, it would be useful to understand better the corrections to the supergravity results originating from the additional presence of winding modes and of $T$-duality in the case of superstrings compactified on smooth manifolds, and those originating from additional non-perturbative states and generalized dualities in compatifications with singularities or localized defects.

Here we would like to make a first step in the direction of more systematic investigations, by considering the simplest higher-dimensional supergravity models, pure $d=5$ supergravities, and by computing their full four-dimensional effective potential $V_1$ after complete supersymmetry breaking by Scherk--Schwarz compactifications. 
We will express $V_1$ as a function of the radial modulus, the only relevant background field in the case under consideration, and of the twist parameters. 
To achieve maximum control on the question we focus on, we will deliberately stay away from  field and string theory models that can be more realistic but require localized defects, such as for example ${\mathbb Z}_2$ orbifold fixed points. 
We will discuss the precise correspondence between the finite effective potential $V_1$ in the full compactified theory and its counterpart $V_{1,red}$ in the reduced theory (divergent for $N=2,4$ and finite for $N=6,8$).
In particular, we will confirm that for completely broken $N=8$ supergravity $V_1$, although corrected with respect to $V_{1,red}$ after accounting for the Kaluza--Klein states,  remains finite and negative definite. 
We will conclude by commenting on the reasons for the qualitatively similar behaviour of $V_1$ and $V_{1,red}$ (with a cut-off $\Lambda \sim 1/R$ when divergent), and on more complicated situations, left for future work, where the study of the full compactified theory could lead to qualitatively new results. 

% section introduction (end)

%\newpage

\section{Scherk--Schwarz compactifications on $S^1$}
\label{sec:sss1}

For Scherk--Schwarz compactifications on a circle, the $d=4$ one-loop effective potential is
\begin{equation}
V_1 = \frac{1}{2} \, 
\int \frac{d^4p}{(2 \pi)^4} \, 
\sum_{n=- \infty}^{+ \infty} \, 
\sum_i \, 
(-1)^{2 J_\alpha} \,
(2 J_\alpha + 1) \,
\log \left( p^2 + m_{n,\alpha}^2 \right)  \, ,
\label{V1gen}
\end{equation}
where the index $\alpha$  runs over the finite number of independent $d=4$ modes of spin $J_\alpha$, with $(2 J_\alpha+1)$ degrees of freedom (dof) and field-dependent mass 
\begin{equation}
m_{n,\alpha}^2 =
\frac{(n+s_\alpha)^2}{R^2}
\qquad
\qquad
\left(
n = 0, \pm1, \pm2, \ldots
\right) 
\label{mass}
\end{equation}
at each of the infinite Kaluza--Klein (KK) levels $n \in \mathbf{Z}$. 
Here $R$ is the physical field-dependent radius of the circle, expressed in units of the reduced Planck mass $M_P$, and the shifts $s_\alpha$ in the mass formula (\ref{mass}) are determined by the Scherk--Schwarz twists.
Since supersymmetry breaking is non-local in the compact dimension and supersymmetric $d=4$ Minkowski vacua are perturbatively stable, $V_1$ is finite \cite{Rohm:1983aq}. 
The explicit calculation can be performed, for example, by adapting the results of  \cite{Delgado:1998qr}, where an explicit cutoff was used at intermediate steps, or those of \cite{Ponton:2001hq}, where zeta-function regularization techniques were used.
The outcome is
\begin{equation}
V_1 = - \frac{3 }{128 \, \pi^6 \, R^4}  \,
\sum_\alpha \, 
(-1)^{2 J_\alpha} \,
(2 J_\alpha + 1) \,
\left[ {\rm Li_5} (e^{- 2 \pi \, i \, s_\alpha}) +  {\rm Li_5} (e^{2 \pi \, i \, s_\alpha})  \right] 
\, ,
\label{V1res}
\end{equation}
where ${\rm Li_n} (x)=\sum_{k=1}^{\infty} x^k/k^n$ are the polylogarithm functions and ${\rm Li_5}(1)=\zeta(5)\simeq 1.037$.
Eq.~(\ref{V1res}) can be generalised to include five-dimensional supersymmetric contributions to the masses, but we won't need such generalisation here. 

It is interesting to understand in some detail the relation between $V_1$ and the potential $V_{1,red}$ computed in the reduced $d=4$ theory.
The latter has an equal and finite number of bosonic and fermionic degrees of freedom, with masses given by Eq.~(\ref{mass}) for $n=0$.
We expect such an effective theory to make sense for $|s_\alpha | \ll 1$, leading to $m_{0,\alpha}^2 \ll (1/R^2)$, with an effective cutoff $\Lambda \sim 1/R$. 
We start from the textbook formula
\begin{equation}
V_{1,red} =
\frac{1}{32 \, \pi^2} \, {\rm Str} \, {\cal M}_0^2 \, \Lambda^2
+
\frac{1}{64 \, \pi^2} \, {\rm Str} \, {\cal M}_0^4 \, \log \frac{{\cal M}_0^2}{\Lambda^2}
\, ,
\label{V4d}
\end{equation}
where
\begin{equation}
 {\rm Str} \, {\cal M}_n^p 
 \equiv
 \sum_\alpha \, 
(-1)^{2 J_\alpha} \,
(2 J_\alpha + 1) \,
m_{n,\alpha}^p
\, .
\label{strdef}
\end{equation}
We observe that, for $|s_\alpha| \ll 1$, the term in square brackets in Eq.~(\ref{V1res}) can be expanded as
\begin{eqnarray}
\left[ {\rm Li_5} (e^{- 2 \pi \, i \, s_\alpha}) +  {\rm Li_5} (e^{2 \pi \, i \, s_\alpha})  \right] 
& = &
2 \, \zeta(5) 
- 4 \pi^2 \, \zeta(3) \, s_\alpha^2
+ \frac{\pi^4}{3} \left[ \frac{25}{3} - 4 \, \log (2 \pi) \right] \, s_\alpha^4
\nonumber \\ & - & 
\frac{2}{3} \, \pi^4 \, s_\alpha^4 \, \log s_\alpha^2
+ \ldots
\, .
\label{V1exp}
\end{eqnarray}
where the first term $2 \, \zeta(5)$ can be ignored, since it does not survive the weighted sum over $\alpha$, equivalent to a supertrace, and  the dots stand for terms of order $s_\alpha^6$ and higher. 
In the reduced theory, ${\rm Str} \, {\cal M}_0^2 = {\rm Str} \, s_\alpha^2 /R^2$. 
Therefore, if $ {\rm Str} \, s_\alpha^2 \ne 0$, the leading contributions to the quadratically divergent $V_{1,red}$ and to the finite $V_1$ are those of order $s_\alpha^2$. 
Equating the two, we find an effective cutoff for the reduced theory
$\Lambda = \sqrt{3 \, \zeta(3)}/(\pi \, R) \simeq 0.6/R$,
in agreement with the expectations.
Suppose now that   ${\rm Str} \, {\cal M}_0^2 = {\rm Str} \, s_\alpha^2/R^2 = 0$ but   ${\rm Str} \, {\cal M}_0^4 = {\rm Str} \, s_\alpha^4/R^4 \ne 0$.  
Then the leading contributions to the logarithmically divergent $V_{1,red}$ and to the finite $V_1$ are those of order $s_\alpha^4$. 
Equating the two, we find an effective cutoff for the reduced theory
$\Lambda = e^{25/3}/(16 \, \pi^4 \, R) \simeq 2.7/R$,
in agreement with the expectations.
Finally, we can consider the case in which both ${\rm Str} \, {\cal M}_0^2 = {\rm Str} \, s_\alpha^2/R^2 = 0$ and  ${\rm Str} \, {\cal M}_0^4 = {\rm Str} \, s_\alpha^4/R^4 = 0$.
In such a case, $V_{1,red}$ and $V_1$ are both finite, with    
$V_{1,red} 
=  {\rm Str} ( {\cal M}_0^4 \, \log {\cal M}_0^2 )/ (64 \, \pi^2)
=  {\rm Str} ( s_\alpha^4 \, \log s_\alpha^2 )/ (64 \, \pi^2 \, R^4)
$
and $V_1$ receiving contributions only from the last line of Eq.~(\ref{V1exp}), therefore coinciding with $V_{1,red}$ up to corrections of order $s_\alpha^6$ or higher.

\subsection{Pure ${\cal N}=2$, $d=5$ supergravity}

This model is described in \cite{tesiwulzer}, in terms of a single twist $a$, which can only break both supersymmetries at once.
For any $n$ there are 8 bosonic dof with $s_\alpha=0$ and $4 \times 2=8$ fermionic dof with $s_\alpha= \pm a$.
For $a=0$ (mod.1) supersymmetry is unbroken: the massless modes are 2 real spin-0 fields, 2 real spin-1 fields, the spin-2 graviton, two Majorana spin-1/2 fermions and two Majorana spin-3/2 gravitinos. 
For any~$a$, the massive modes at each KK level are a vector, the graviton and two gravitinos.   
Then
%
%\begin{equation}
%V_1 = \frac{3}{16 \, \pi^6 \, R^4}  \,
%\left[ {\rm Li_5} (e^{- 2 \pi \, i \, a}) +  {\rm Li_5} (e^{- 2 \pi \, i \, a}) - 2 \, \zeta(5) \right] 
%\, .
%\label{V1c}
%\end{equation}
%
$V_1$ in (\ref{V1res}) is negative semidefinite, vanishes only for unbroken supersymmetry and for fixed $R$ is minimized by $a=1/2$ (mod.1), where ${\rm Li_5} (-1) = -(15/16) \, \zeta(5) \simeq -0.972$. 
Notice also that  ${\rm Str} \, {\cal M}_0^2 =  {\rm Str} \, {\cal M}_n^2  =  {\rm Str} \, s_\alpha^2 /R^2 = - 8 a^2/R^2<0$: the quadratic supertrace in the reduced theory is identical to that at any fixed Kaluza--Klein level $n \ne 0$ in the compactified theory.  

\subsection{Pure ${\cal N}=4$, $d=5$ supergravity}

The model is described in \cite{Villadoro:2004ci}, in terms of two independent twists $a_i$ $(i=1,2)$, each of which can independently break two of the four supersymmetries. 
For any $n$, there are 24 bosonic and 24 fermionic dof. 
Of the bosonic ones, 12 have $s_\alpha=0$ (for $n \ne 0$, 1 spin-2, 2 real spin-1 and 1 real spin-0; for $n=0$, 1 spin-2, 3 real spin-1 and 4 real spin-0), $3 \times 4 =12$ have $s_\alpha=\pm a_1 \pm a_2$ (4 real spin-1). 
Of the fermionic ones, $6 \times 2=12$ have $s_\alpha= \pm a_1$ (2 Majorana spin-3/2 and 2 Majorana spin-1/2)  and $6 \times 2=12$ have $s_\alpha= \pm a_2$ (2 Majorana spin-3/2 and 2 Majorana spin-1/2).     
%Then
%
%$$
%V_1 = \frac{9}{64 \, \pi^6 \, R^4}  \,
%\left\{
%2 \left[ {\rm Li_5} (e^{- 2 \pi \, i \, a_1}) +  {\rm Li_5} (e^{2 \pi \, i \, a_1})  \right]
%+
%2 \left[ {\rm Li_5} (e^{- 2 \pi \, i \, a_2}) +  {\rm Li_5} (e^{2 \pi \, i \, a_2})  \right]
%- 4 \, \zeta(5)
%\right.
%$$
%\begin{equation}
%\left.
%-
%\left[ {\rm Li_5} (e^{- 2 \pi \, i \, (a_1-a_2)}) +  {\rm Li_5} (e^{2 \pi \, i \, (a_1-a_2)})  \right]
%-
%\left[ {\rm Li_5} (e^{- 2 \pi \, i \, (a_1+a_2)}) +  {\rm Li_5} (e^{2 \pi \, i \, (a_1+a_2)})  \right]
% \right\}
%\, .
%\label{V1c}
%\end{equation}
%
Also in this case $V_1$ in (\ref{V1res}) is negative semidefinite, vanishes only in the case of some unbroken supersymmetry and for fixed $R$ is minimized by $a_1=a_2=1/2$ (mod.1). 
Notice also that ${\rm Str} \, {\cal M}_0^2 =  {\rm Str} \, {\cal M}_n^2 =  0$, as expected for spontaneously broken ${\cal N}=4$ supersymmetry, and ${\rm Str} \, {\cal M}_0^4 =   {\rm Str} \, {\cal M}_n^4 =  {\rm Str} \, s_\alpha^4/R^4 = 72 \, a_1^2 \, a_2^2/R^4>0$:
both the quadratic and the quartic supertrace are the same in the reduced theory and at any fixed Kaluza--Klein level $n \ne 0$ in the compactified theory.  

\subsection{${\cal N}=6$, $d=5$ supergravity}

The model is described in \cite{Villadoro:2004ec}, in terms of three independent twists $a_i$ $(i=1,2,3)$, each of which can independently break two of the six supersymmetries. 
For any $n$, there are 64 bosonic and 64 fermionic dof. 
Of the bosonic ones:
16 have $s_\alpha=0$; 
4 each have $s_\alpha=\pm a_i \pm a_j$ $(i<j)$, for a total of $4 \times 4 \times 3 = 48$. 
Of the fermionic ones:
4 each have $s_\alpha= \pm a_i$ $(i=1,2,3)$, for a total of $4 \times 4 \times 3=48$; 
2 each have $s_\alpha= \pm a_i \pm a_j \pm a_k$, $(i<j<k)$, for a total of $2 \times 8 =16$. 
Also in this case $V_1$ in (\ref{V1res}) is negative semidefinite, vanishes only in the case of some unbroken supersymmetry and for fixed $R$ is minimised by $a_1=a_2=a_3=1/2$ (mod.1), assuming the value
\begin{equation}
V_1 =
- \frac{93 \, \zeta(5)}{16 \, \pi^6 \, R^4} 
\simeq
- \frac{0.0063}{R^4}
\, .
\label{V6f}
\end{equation}
In the reduced theory, it is known that ${\rm Str} \, {\cal M}_0^2 = {\rm Str} \, {\cal M}_0^4 = 0$ and 
${\rm Str} \, {\cal M}_0^6 =  {\rm Str} \, s_\alpha^6/R^6 = - 1440 \, a_1^2 \, a_2^2 \, a_3^2 / R^6$.
We find that the same result holds true at any fixed Kaluza--Klein level $n \ne 0$ in the compactified theory:
${\rm Str} \, {\cal M}_n^2 = {\rm Str} \, {\cal M}_n^4 = 0$ and 
${\rm Str} \, {\cal M}_n^6 =  {\rm Str} \, s_\alpha^6/R^6 = -1440 \, a_1^2 \, a_2^2 \, a_3^2 / R^6$.
In the limit $|a_i| \ll 1$  (for all $i=1,2,3$), the contributions of the KK modes become negligible and $V_1 \simeq V_{1,red}$, up to  corrections of order $s_\alpha^6$ and higher.
Instead, for  $a_1=a_2=a_3=1/2$, we find:
\begin{equation}
V_{1,red} = 
- \frac{3 (32 \, \log 2 - 27 \, \log 3)}{128 \, \pi^2 \, R^4} 
\simeq
- \frac{0.018}{R^4}
\, ,
\label{V6r}
\end{equation}
three times larger than $V_1$ computed for the same twists in the full compactified theory.

\subsection{${\cal N}=8$, $d=5$ supergravity}

The model is described in \cite{Cremmer:1979uq}, in terms of four independent twists $a_i$ $(i=1,2,3,4)$, each of which can independently break two of the eight supersymmetries. 
For any $n$, there are 128 bosonic and 128 fermionic dof. 
Of the bosonic ones:
16 have $s_\alpha=0$; 
4 each have $s_\alpha=\pm a_i \pm a_j$ $(i<j)$, for a total of $4 \times 4 \times 6=96$;
1 each have $s_\alpha = \pm a_1 \pm a_2 \pm a_3 \pm a_4$, for a total of 16. 
Of the fermionic ones:
8 each have $s_\alpha= \pm a_i$ $(i=1,2,3,4)$, for a total of $8 \times 2 \times 4=64$; 
2 each have $s_\alpha= \pm a_i \pm a_j \pm a_k$, $(i<j<k)$, for a total of $2 \times 8 \times 4=64$. 
Also in this case $V_1$ in (\ref{V1res}) is negative semidefinite, vanishes only in the case of some unbroken supersymmetry and for fixed $R$ is minimised by $a_1=a_2=a_3=a_4=1/2$ (mod.1), assuming the value
\begin{equation}
V_1 =
- \frac{93 \, \zeta(5)}{8 \, \pi^6 \, R^4} 
\simeq
- \frac{0.0125}{R^4}
\, .
\label{V8f}
\end{equation}
In the reduced theory, it is known that ${\rm Str} \, {\cal M}_0^2 = {\rm Str} \, {\cal M}_0^4 = {\rm Str} \, {\cal M}_0^6 = 0$ and 
${\rm Str} \, {\cal M}_0^8 =  {\rm Str} \, s_\alpha^8/R^8 = 40320 \, a_1^2 \, a_2^2 \, a_3^2 \, a_4^2 / R^8$.
We find that the same result holds true at any fixed Kaluza--Klein level $n \ne 0$ in the compactified theory:
${\rm Str} \, {\cal M}_n^2 = {\rm Str} \, {\cal M}_n^4 = {\rm Str} \, {\cal M}_n^6 = 0$ and 
${\rm Str} \, {\cal M}_n^8 =  {\rm Str} \, s_\alpha^8/R^8 = 40320 \, a_1^2 \, a_2^2 \, a_3^2 \, a_4^2 / R^8$.
In the limit $|a_i| \ll 1$  (for all $i=1,2,3,4$), the contributions of the KK modes become negligible and $V_1 \simeq V_{1,red}$, up to  corrections of order $s_\alpha^8$ and higher.
Instead, for  $a_1=a_2=a_3=a_4=1/2$, we find:
\begin{equation}
V_{1,red} = 
- \frac{3 (40 \, \log 2 - 27 \, \log 3)}{32 \, \pi^2 \, R^4} 
\simeq
- \frac{0.0184}{R^4}
\, ,
\label{V8r}
\end{equation}
not too far from $V_1$ computed for the same twists in the full compactified theory.

\section{Conclusions and outlook} % (fold)
\label{sec:cando}

In the case of a single extra dimension compactified on a circle, considered in this paper, the Scherk--Schwarz twist leading to supersymmetry breaking corresponds to a global continuous R-symmetry of the higher-dimensional action. 
We can smoothly take the limit of small twist parameters, $s_{\alpha} \ll 1$, and continuously connect the full effective potential $V_1$ of the compactified theory with the effective potential $V_{1,red}$ of the reduced theory.
Moreover, the tree-level spectrum depends on a single modulus $R$. 
It is then not surprising that $V_1$ and $V_{1,red}$ have a very similar behaviour, and that in the $N=8$ case the disappointing result obtained for $V_{1,red}$ is now corrected but qualitatively confirmed also for $V_1$. 

However, when the compactification involves three or more internal dimensions we can use general coordinate transformations to twist boundary conditions.
After a field redefinition, twisted boundary conditions can be replaced by a VEV for the spin connection, which leads to observable effects because the compact space is non-simply connected, in analogy with gauge symmetry breaking by Wilson lines.
This kind of Scherk--Schwarz compactifications are also known as twisted tori compactifications, since the boundary conditions act on the geometry of the internal manifold, modifying its structure.
In such cases, the twist parameters are quantized and the reduced theory can either be a mere consistent truncation or a genuine effective theory \cite{DallAgata:2005zlf, Grana:2013ila}, depending on whether the compactification manifold is homogeneous or not. 
In the first case, the three-dimensional manifold is still a torus and supersymmetry appears to be broken in the reduced truncated theory but remains unbroken in the full compactified theory. 
In such a case, we expect $V_1=0$, i.e. perturbative stability for the Minkowski vacuum of the full higher-dimensional theory, which is indeed supersymmetric, although $V_{1,red}$ could have a completely different qualitative behaviour.   
In the second case, the relation bewteen $V_1$ and $V_{1,red}$ is an open problem that deserves further investigations \cite{GDAFZ2024}.
The tree-level field-dependent spectrum of the compactified theory may have a non-trivial structure, with masses depending on several moduli and the absence of tachyons guaranteed only in certain regions of the moduli space. 
As a consequence, also the full 1-loop potential $V_1$, obtained by summing over the contributions of all Kaluza--Klein states, could exhibit significantly different qualitative features from $V_{1,red}$ computed in the truncated theory.
 
The simplest scenario we can consider is three extra dimensions, with the internal manifold given by a twisted torus whose geometry is that of the quotient  ${\mathbb E}_2/\Gamma$ of the two-dimensional Euclidean group ${\mathbb E}_2$  by one of the 17 inequivalent discrete subgroups $\Gamma \subset {\mathbb E}_2$ called wallpaper groups \cite{CoxeterMoser}, with the exclusion of the trivial case with translations only, which gives rise to a torus and does not break supersymmetry in the full compactified theory.
The main obstacle in computing $V_1$ is finding a complete set of harmonic functions on ${\mathbb E}_2/\Gamma$, a prerequisite for computing the full Kaluza-Klein spectrum and eventually $V_1$, 
After obtaining a basis for the scalar harmonics, we could adapt the powerful techniques recently developed in \cite{Malek:2019eaz,Malek:2020yue} to compute the full Kaluza-Klein spectra. 
In any case, Scherk--Schwarz geometrical fluxes are the only `flat gaugings' that do not require D-branes and/or orientifolds to comply with known no-go theorems for Minkowski vacua of higher dimensional theories \cite{deWit:1986mwo,Maldacena:2000mw}: they are therefore the obvious next step to explore, and probably the last one where stringy effects can be consistently neglected.

\acknowledgments
This work was supported in part by the Italian MUR Departments of Excellence grant 2023-2027 ``Quantum Frontiers'' and by the MUR-PRIN contract 2022YZ5BA2 - ``Effective quantum gravity''.
%\newpage

% section the_models (end)

%
%\acknowledgments
%We thank \ldots for discussions. G.D.A. was supported in part by \ldots and F.Z. was supported in part by \ldots.
%

\end{document}